\documentclass[pra,aps,twocolumn,superscriptaddress,groupedaddress,showpacs]{revtex4}
\usepackage{psfig}
\usepackage{epsf}

\begin{document}

\title{Kerr nonlinear coupler and entanglement}

\author{Wies\a law Leo\'nski}

\address{Institute of Physics, Adam Mickiewicz University,
ul. Umultowska 85, 61-614 Pozna\'n, Poland}

\author{Adam Miranowicz}

\address{Institute of Physics, Adam Mickiewicz University,
ul. Umultowska 85, 61-614 Pozna\'n, Poland}

\address{CREST Research Team for Interacting Carrier
Electronics, Graduate University for Advanced Studies (SOKENDAI),
Hayama, Kanagawa 240-0193, Japan}

\begin{abstract}
We discuss a model comprising two coupled nonlinear oscillators
(Kerr-like nonlinear coupler) with one of them pumped by an
external coherent excitation. Applying the method of nonlinear
quantum scissors we show that the quantum evolution of the coupler
can be closed within a finite set of $n$-photon Fock states.
Moreover, we show that the system is able to generate Bell-like
states and, as a consequence, the coupler discussed behaves as a
two-qubit system.  We also analyze the effects of dissipation on
entanglement of formation parametrized by concurrence.
\end{abstract}


\maketitle

\section{Introduction}
Quantum entanglement seems to be one of the most striking
phenomena of quantum physics. It is not only one of the most
fundamental concepts of quantum information theory, but also
allows investigation of many features of nonlocal properties of
quantum systems \cite{Ein35}.  Various aspects of the entanglement
and its generation have been discussed in numerous papers,
especially from the point of view of quantum information
applications including quantum key distribution \cite{Eke91},
superdense coding \cite{Ben92}, quantum teleportation
\cite{Ben93}, fast quantum computations \cite{Sho94,Gro97},
entanglement-assisted communication \cite{Ben97} or broadcasting
of entanglement \cite{Buz97}.

In this paper we shall concentrate on the dynamics of the {\em
Kerr nonlinear coupler} and its ability to produce quantum
entangled states. Since the pioneering works of Jensen
\cite{Jen82} and Maier \cite{Mai82}, the nonlinear couplers
attract an increasing interest \cite{Che96}--\cite{Gry01} (for
reviews see \cite{Per00,Fiu01}). As shown in classical
\cite{Jen82,Mai82} and quantum \cite{Che96} models, the Kerr
couplers can exhibit self-trapping, self-modulation and
self-switching of the energy of the coupled modes. These phenomena
have potential applications in optical communications as, e.g.,
intensity-dependent routing switches. Among various other quantum
statistical properties, it has been shown that the Kerr couplers
can be a source of sub-Poissonian and squeezed light
\cite{Per95}--\cite{Ari00}. Another group of papers concerns the
correspondences between the quantum and classical dynamics of such
systems \cite{Che96} and their chaotic dynamics, including
synchronization effects \cite{Gry01}.

Quantum optical systems based on Kerr nonlinearity have been
applied for various quantum information purposes including
entanglement purification \cite{Dua00}, complete quantum
teleportation \cite{Vit00}, or realization of qubit phase gates
\cite{Ott03}.  Here, we present another simple quantum information
application Kerr nonlinearities, namely for generation of
entangled optical qubits from classical light.

We are interested here in a simple model comprising two quantum
nonlinear oscillators located inside one cavity. These oscillators
are linearly coupled to each other, while one of the oscillators
is excited by an external coherent field of a constant amplitude.
For this model we shall answer the questions, whether it is
possible to close the dynamics of the excited nonlinear coupler
within a finite set of $n$-photon states and, which is the main
subject of this paper, whether nonlinear excited coupler can be a
source of maximally entangled (ME) states.

\section{The model and solutions}

\begin{figure}
\hspace*{0.5cm}{\psfig{figure=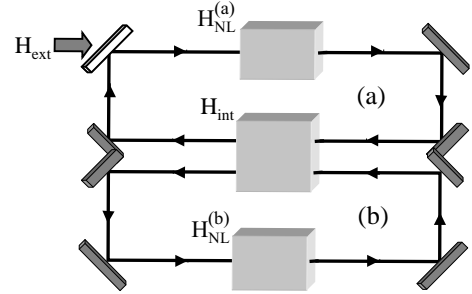,width=7cm}}\vspace{-5mm}
\caption{Scheme of a pumped nonlinear coupler, described by
Hamiltonian (\ref{8}), implemented by two ring cavities ($a$ and
$b$) filled with Kerr media, where cavity $a$ is being excited by
a single-mode external classical field.}
\end{figure}
The model of the Kerr nonlinear coupler discussed here contains
two nonlinear oscillators linearly coupled to each other and,
additionally, one of them is coupled to an external coherent field
as presented in figure 1. We assume, that this excitation is
linear and has a constant amplitude. This system can be described
by the following Hamiltonian:
\begin{equation}
\hat{H}=\hat{H}_{NL}+\hat{H}_{int}+\hat{H}_{ext} \label{8}
\end{equation}
where
\begin{eqnarray}
\hat{H}_{NL}&\equiv&\hat{H}^{(a)}_{NL}+\hat{H}^{(b)}_{NL}
=\frac{\chi_a}{2}(\hat{a}^\dagger )^2\hat{a}^2+
\frac{\chi_b}{2}(\hat{b}^\dagger )^2\hat{b}^2, \label{9a}\\
\hat{H}_{int}&=&\epsilon\hat{a}^\dagger\hat{b} +
\epsilon^*\hat{a}\hat{b}^\dagger, \label{9b}\\
\hat{H}_{ext}&=&\alpha\hat{a}^\dagger+\alpha^*\hat{a}. \label{9c}
\end{eqnarray}
We see that $\hat{H}_{NL}$ describes nonlinear oscillators,
$\hat{H}_{int}$ corresponds to an internal coupling, whereas the
term $\hat{H}_{ext}$ describes a linear coupling between the
external field and the mode of the field inside our cavity
corresponding  to the oscillator $a$. The parameters $\chi_a $ and
$\chi_b$ are nonlinearity constants of the oscillators $a$ and
$b$, respectively, $\epsilon$ describes the strength of the
oscillator -- oscillator coupling, whereas $\alpha$ is a strength
of the external excitation of the oscillator $a$.  It is worth
noting that our Hamiltonian $\hat{H}_{int}$ does not include
nonlinear coupling proportional to
$\hat{b}^\dagger\hat{b}\hat{a}^\dagger\hat{a}$ but only the linear
one, described by (\ref{9b}). Nevertheless, the same Hamiltonian
$\hat{H}_{NL}+\hat{H}_{int}$ as ours was used, e.g., by Bernstein
\cite{Ber93} and Chefles and Barnett \cite{Che96} to describe the
nonlinear coupler.

In the first part of our analysis we neglect damping processes in
our model, thus the system evolution can be described by a
time-dependent wave-function. This function can be written in the
$n$-photon Fock basis as:
\begin{equation}
|\psi (t)\rangle  =\sum_{n,m=0}^{\infty}c_{n,m}(t)| n\rangle _a|
m\rangle _b \label{2}
\end{equation}
where $c_{n,m}(t)$ is a complex probability amplitude of finding
our system in the $n$-photon and $m$-photon states for the mode
$a$ and $b$, respectively.

We have included here an external coupling and therefore, the
energy inside the cavity is not conserved. As a consequence, we
can expect that in the evolution of the system many of the states
corresponding to great number of photons will be involved.
However, we can overcome this difficulty by applying the {\em
nonlinear quantum scissors} method discussed in \cite{Leo97} (for
the discussion concerning quantum states defined in
finite-dimensional Hilbert spaces and the methods of their
generation see the review papers \cite{Mir01,Leo01} and the
references cited therein). Namely, it is seen from the form of
$\hat{H}_{NL}$ that this Hamiltonian produces degenerate levels of
the energy equal to zero, corresponding to the following four
states: $|0\rangle _a|0\rangle _b$, $|1\rangle _a|0\rangle _b$,
$|0\rangle _a|1\rangle _b$ and $|1\rangle _a|1\rangle _b$.
Moreover, all couplings discussed here have constant envelopes,
and similarly as in \cite{Leo97}, we assume that they are weak.
Therefore, we can treat transitions within the mentioned set of
the states as that of resonant nature. The evolution of discussed
system is closed within the set of these four states and
interactions with other states can be neglected in our
approximation. Thus, the wave-function describing our model can be
written in the form:
\begin{eqnarray}
|\psi (t)\rangle  &=& c_{0,0}(t)| 0\rangle _a| 0\rangle
_b+c_{1,0}(t)| 1\rangle _a| 0\rangle _b \nonumber\\&& +c_{0,1}(t)|
0\rangle _a| 1\rangle _b+c_{1,1}(t)| 1\rangle _a| 1\rangle
_b\,\,\, \label{10}
\end{eqnarray}
and hence, the equations of motion for the system are:
\begin{eqnarray}
i\frac{d}{dt}c_{0,0}&=&\alpha^*c_{1,0},\nonumber\\
i\frac{d}{dt}c_{1,0}&=&\epsilon\, c_{0,1}+\alpha\, c_{0,0},\nonumber\\
i\frac{d}{dt}c_{0,1}&=&\epsilon^*c_{1,0}+\alpha^*c_{1,1},\nonumber\\
i\frac{d}{dt}c_{1,1}&=&\alpha\, c_{0,1} . \label{11}
\end{eqnarray}
To solve these equations we need to find roots of the forth-order
polynomial, which leads to a very complicated and unreadable form
of final formulas. Therefore, although it is formally possible, we
shall not write the analytical solutions in their most general
form and we will restrict our considerations to the case of real
$\alpha=\epsilon$. Moreover, we assume that for the time $t=0$ the
both oscillators were in vacuum states, i.e.,
\begin{eqnarray}
|\psi (t=0)\rangle  =| 0\rangle _a| 0\rangle _b . \label{12}
\end{eqnarray}
Then we get the following solutions for the probability amplitudes
$c_{i,j}$ ($i,j=0,1)$:
\begin{eqnarray}
c_{0,0}(t) &=&\cos (xt) \cos (yt)
+\frac{1}{\sqrt{5}}\sin (xt) \sin (yt), \nonumber \\
c_{1,0}(t) &=&-i\frac{2}{\sqrt{5}}\cos (xt)\sin (yt) ,
\nonumber\\
c_{0,1}(t) &=&-\frac{2}{\sqrt{5}}\sin (xt)\sin (yt) ,
\nonumber \\
c_{1,1}(t) &=& i[\frac{1}{\sqrt{5}} \cos (xt) \sin (yt)  -\sin
(xt) \cos (yt) ] \label{13}
\end{eqnarray}
where $x=\alpha/2$ and $y=\sqrt{5}x$.  The solution (\ref{13}) is
valid under the condition $\chi_j \gg \epsilon=\alpha$ ($j=a,b$),
which implies that it is apparently independent of nonlinearities
$\chi_j$. But it should be stressed that the corresponding
nonlinear Hamiltonian $\hat{H}_{NL}$, given by (\ref{9a}), is
responsible for the truncation of the infinite-dimensional state
to the finite superposition, given by (\ref{10}). Otherwise, if
$\chi_j$ were not much stronger than $\epsilon$ and $\alpha$, the
state generated would not be truncated to the finite superposition
(\ref{10}) and the probability amplitudes $c_{n,m}(t)$ would
depend explicitly on nonlinearities $\chi_j$.

To check our solutions we can calculate the probability amplitudes
numerically in  a basis expanded to the states corresponding to
greater number of photons than discussed here (for the model
discussed our considerations are restricted by the resonances to
the one-photon and vacuum states only). We perform the
calculations following the method discussed in \cite{Leo01}, and
first we construct the unitary evolution operator $\hat{U}$
applying the full Hamiltonian shown in (\ref{8}):
\begin{eqnarray}
\hat{U}&=&\exp\left(-i\hat{H}t\right). \label{14}
\end{eqnarray}
Then we are able to obtain the wave-function $|\psi (t)\rangle $
by acting the operator $\hat{U}$ on the initial state of the
system, and for the case discussed here we have:
 \begin{eqnarray}
|\psi(t)\rangle &=&\hat{U}(t)|0\rangle _a|0\rangle _b .\label{15}
\end{eqnarray}
\begin{figure}
\psfig{figure=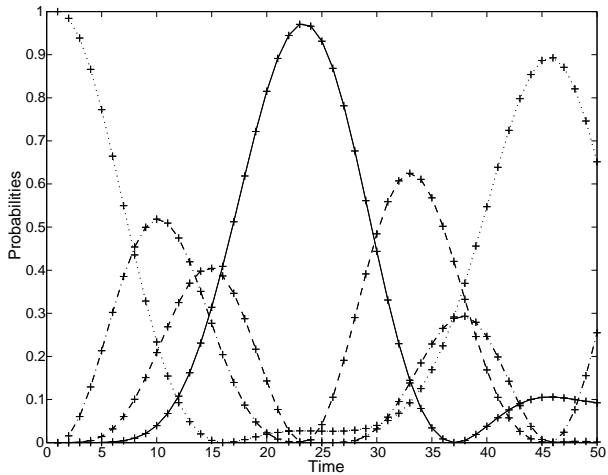,width=8cm}
\caption{\label{fig:epsart1} Probabilities for finding the coupler
in the $|0\rangle _a|0\rangle _b$ (dotted curve), $|1\rangle
_a|0\rangle _b$ (dashed-dotted curve)$|0\rangle _a|1\rangle _b$
(dashed curve) and $|1\rangle _a|1\rangle _b$ (solid curve) states
from the analytical results and their numerical counterparts
(cross marks). The nonlinearity coefficients $\chi_a=\chi_b=25$
and the coupling strengths $\epsilon=\alpha=\pi/25$.}
\end{figure}
\begin{figure}
 \vspace*{-4mm} \hspace*{-3mm}
 \epsfxsize=4cm\epsfbox{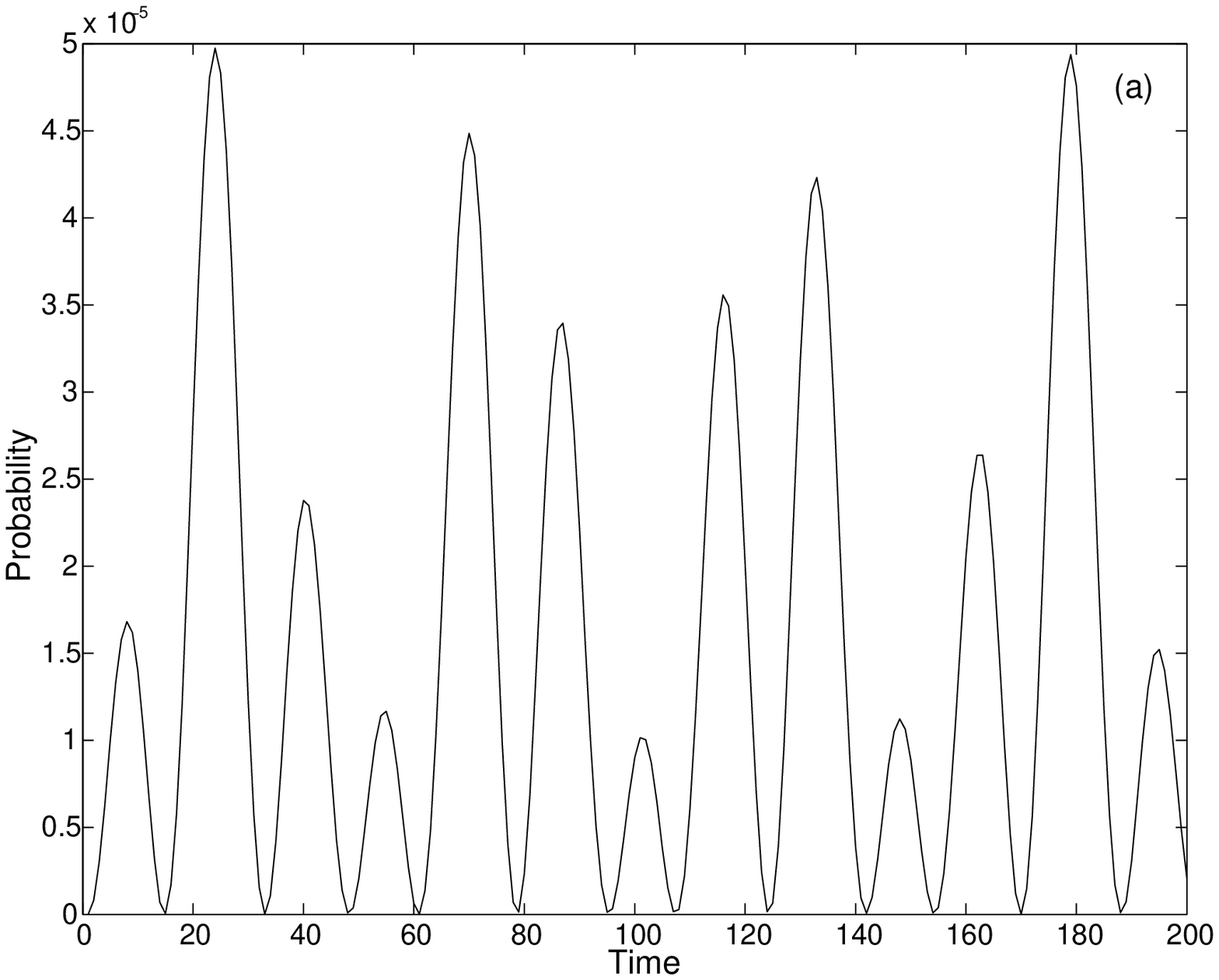}
 \epsfxsize=4cm\epsfbox{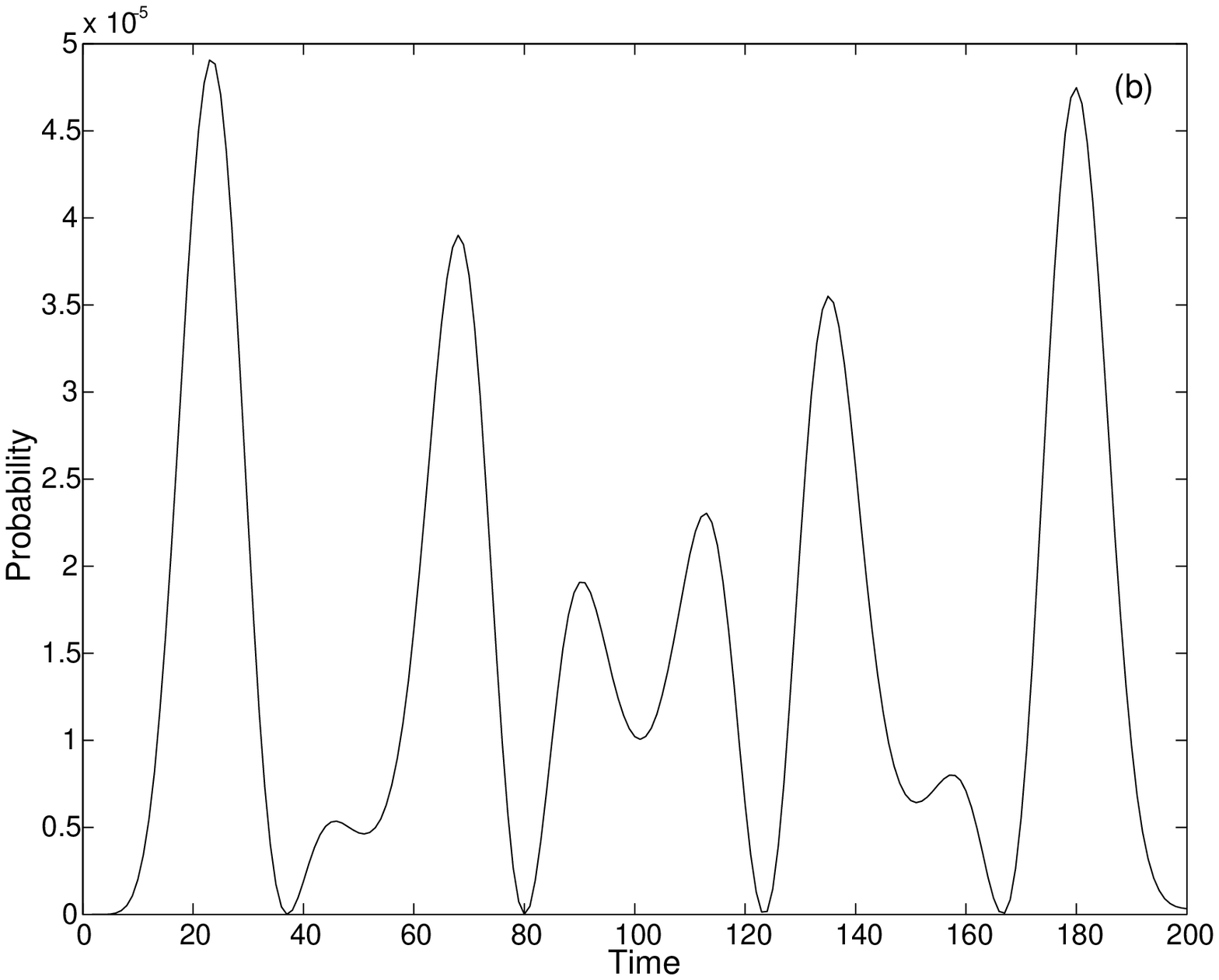}
\caption{\label{fig:epsart2} The same as in figure 2 but for the
states: (a) $|0\rangle _a|2\rangle _b$ and (b) $|1\rangle
_a|2\rangle _b$.}
\end{figure}
Figure 2 shows both analytical and numerical results of our
calculations. We see very good agreement between these two
methods, so the model based on the resonances works very well.
Moreover, from the numerical results for the probabilities
corresponding to the states $|0\rangle _a|2\rangle _b$ (figure 3a)
and $|1\rangle _a|2\rangle _b$ (figure 3b), we see that the states
corresponding to the numbers of photons higher than one are
practically unpopulated. It is worth mentioning that our numerical
calculations have been performed in the $m$-dimensional Fock
basis, where $m\simeq 20$ for each subspace associated with a
single mode of the field.

\section{Coupler and entanglement}

The time-evolution of the probability amplitudes can give some
information concerning entanglement in our system too. For
instance, if we see in a figure that probability corresponding to
one of the discussed states is equal to another one and
additionally, the both are equal to $1/2$ ($|c_{i,j}|^2
=|c_{k,l}|^2 =0.5$ for every $i,j,k,l$), we know that the system
generates ME states. Obviously, this method of finding entangled
states is not very accurate, especially for the case when we
should observe and compare various and often rapidly oscillating
probabilities. Therefore, we apply another method convenient for
finding entanglement in the system. Namely we shall express
obtained wave-function in the Bell basis:
\begin{eqnarray}
|\psi\rangle =b_1|B_1\rangle +b_2|B_2\rangle +b_3|B_3\rangle
+b_4|B_4\rangle  \label{16}
\end{eqnarray}
where the states $|B_i\rangle $, $i=1,2,3,4$ are Bell-like states
that can be expressed as functions of discussed here $n$-photon
states (Bell-like states differ from the commonly discussed Bell
states in the existence of the phase factor -- for the case
discussed here, one of the $n$-photon states is multiplied by $i$)
:
\begin{eqnarray}
|B_1\rangle &=&\frac{1}{\sqrt{2}}\left(|1\rangle _a|1\rangle _b
+i|0\rangle _a|0\rangle _b\right),\nonumber\\
|B_2\rangle &=&\frac{1}{\sqrt{2}}\left(|0\rangle _a|0\rangle _b
+i|1\rangle _a|1\rangle _b\right),\nonumber\\
|B_3\rangle &=&\frac{1}{\sqrt{2}}\left(|0\rangle _a|1\rangle _b
-i|1\rangle _a|0\rangle _b\right),\nonumber\\
|B_4\rangle &=&\frac{1}{\sqrt{2}}\left(|1\rangle _a|0\rangle
_b-i|0\rangle _a|1\rangle _b\right). \label{17}
\end{eqnarray}
These states are maximally entangled states, and therefore, for
the cases when $|b_i|^2=1$, $i=1,2,3,4$ also our system evolves
into ME state. Figure 4 shows probabilities corresponding to the
Bell-like states as a function of time. Moreover, all parameters
describing our system are identical to those of figure 2.
\begin{figure}
\psfig{figure=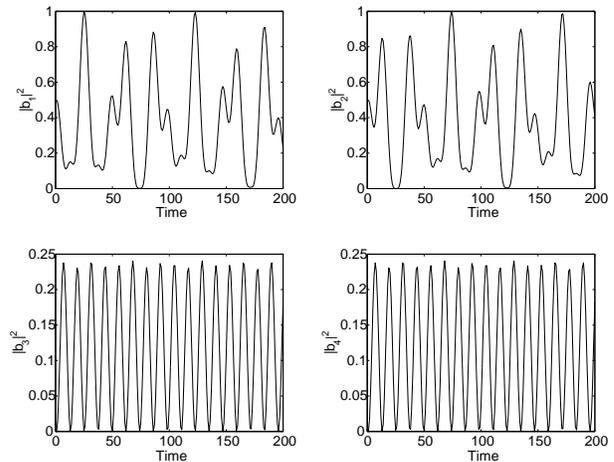,width=8cm}
\caption{\label{fig:epsart4} Probabilities  for finding the
coupler in the Bell-like states.  The nonlinearity coefficients
$\chi_a$ and $\chi_b$, and the coupling strengths $\epsilon$ and
$\alpha$ are identical to those of figures 2--3.}
\end{figure}
We see, that for the time $t\simeq 115$ we get the state
$|B_1\rangle $ and for $t\simeq 80$ the state $|B_2\rangle $ is
generated with high accuracy -- our system becomes maximally
entangled. This entanglement involves the states $|0\rangle
_a|0_b\rangle $ and $|1\rangle _a|1_b\rangle $. Of course, one
should keep in mind that plots in figure 4 are for the
probabilities not for their complex amplitudes and hence, we get
the Bell-like states from (\ref{17}) with some phase factor.
Nevertheless, our states are maximally entangled. Moreover, figure
4 shows that the values of probabilities for the states
$|B_3\rangle $ and $|B_4\rangle $ can reach maximally  $0.8$. As a
consequence, the states $|1\rangle _a|0_b\rangle$ and $|0\rangle
_a|1_b\rangle $ cannot be maximally entangled  for the initial
vacuum states $|0\rangle _a|0_b\rangle$. But generation of
$|B_3\rangle $ and $|B_4\rangle $ would be possible by assuming
that the system is initially in the states $|1\rangle
_a|0_b\rangle$ or $|0\rangle _a|1_b\rangle $.

The Bell-like states (\ref{17}) are maximally entangles, however,
they are not the only entangled states that could be produced by
the system. Therefore, to measure the entanglement degree of the
system we apply the measure that is referred to as the {\em
concurrence}. This quantity proposed by Wootters \cite{Woo98} is
one of the most commonly applied measures of the entanglement. The
concurrence for  two qubit states is defined as
\begin{equation}
{\cal C}=\max \left\{0,
\lambda_1-\lambda_2-\lambda_3-\lambda_4\right\} \label{18}
\end{equation}
where $\lambda_i$ ($i=1,\ldots ,4$) are the square roots of the
eigenvalues of the matrix
\begin{equation}
\tilde{\rho}=\rho\left(\sigma^a_y\otimes\sigma^b_y\right)\rho^*
\left(\sigma^a_y\otimes\sigma^b_y\right) \label{19}
\end{equation}
where $\sigma^{\{a,b\}}_y$ are Pauli matrices defined in subspaces
corresponding to the modes $\{a,b\}$, and the eigenvalues
$\lambda_i$ appearing in (\ref{18}) should be taken in decreasing
order. Concurrence takes values from $0$ to $1$, where for
unentangled states it vanishes, whereas for ME states it is equal
to one.

Damping is the main and unavoidable source of decoherence which
can easily destroy entangled states. Hence, for our results to be
applicable in real physical systems, we present a numerical
analysis of the damping effects on the concurrence. Let us assume
that the leakage of photons from the cavities $a$ and $b$ is
described by the rates $\kappa_a$ and $\kappa_b$, respectively.
Starting from Hamiltonian (\ref{8}) and defining the collapse
operators by $\hat{C}_a=\sqrt{2\kappa_a}\hat{a}$ and
$\hat{C}_b=\sqrt{2\kappa_b}\hat{b}$, we can write the
time-independent Liouvillian in the standard Linblad form
\begin{equation}
\hat{{\cal L}}\hat{\rho}=-i[\hat{H},\hat{\rho}]+\sum_{j=a,b}\left(
\hat{C}_j\hat{\rho} \hat{C}_j^{\dagger}-\frac12(
\hat{C}_j^{\dagger}\hat{C}_j\hat{\rho}
+\hat{\rho}\hat{C}_j^{\dagger}\hat{C}_j)\right). \label{18a}
\end{equation}
The evolution of the density matrix $\hat{\rho}(t)$ in the
dissipative system can be found numerically as a series of complex
exponentials $\exp(\sigma_k t)$ given in terms of the eigenvalues
$\sigma_k$ of the Liouvillian $\hat{{\cal L}}$, given by
(\ref{18a}). \begin{figure}
\psfig{figure=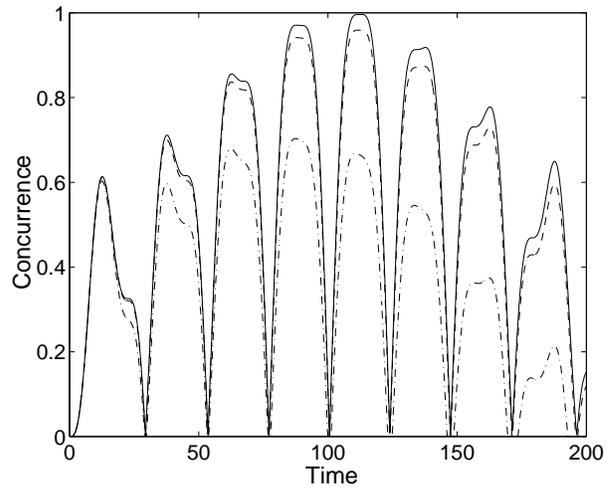,width=8cm}
\caption{\label{fig:epsart5} Concurrence for the excited nonlinear
coupler for various cavity leakage rates $\kappa_a=\kappa_b$ equal
to 0 (solid curve), $10^{-4}$ (dashed curve) and $10^{-3}$
(dot-dashed curve). The quantities $\chi_a$, $\chi_b$, $\epsilon$
and $\alpha$ are identical to those of figure 4.}
\end{figure}

Figure 5 shows the plot of the concurrence evolution for the case
discussed here -- we have single external excitation of the
coupler and we assume that all couplings existing in the system
are weak ($\alpha=\epsilon=\pi /25$) in comparison to Kerr
nonlinearities $\chi_a=\chi_b=25$. Various curves in figure 5
correspond to concurrence evolutions with different dissipation
rates. We see that the varying in time concurrence is modulated by
an oscillation of low frequency. As a consequence, several maxima
appearing here are of various values. Two of them for
dissipation-free evolution (depicted by solid curve), which are
the closest to unity, correspond to the formation of Bell-like
states $|B_1\rangle $ and $|B_2\rangle $ discussed earlier and
shown in figure 4. As a consequence, we can treat our system as a
source of ME states for low dissipation. On the scale of figure 5
and for the chosen coupling parameters, the differences between
the evolution with the leakage rates $\kappa_a=\kappa_b\le
10^{-5}$ and the dissipation-free evolution are invisible.
However, higher leakage rates beyond short time evolution cause
essential deterioration of the concurrence limiting the effective
generation of ME states. Thus, the results of our calculations
indicate that the system discussed is highly sensitive to the
dissipation processes. Even relatively small losses from the
cavity are able to destroy the entanglement. Therefore, we should
assume that we deal here with a very high $Q$ cavity that is
capable to preserve practically the whole radiation field located
inside. However, this assumption is very desirable from our point
of view. For this case the coupler can be weakly excited by
external fields only -- the less photons can escape from the
cavity through the mirror, the smaller number of photons can be
injected inside this way.

\section{Conclusions}

In this paper we have discussed a model of nonlinear coupler
linearly excited by a single-mode coherent field. We have shown
that   the evolution of the system is closed within a finite set
of states and only $|i\rangle _a|j\rangle _b$ ($i,j=0,1$) states
are populated. We have applied here the method used for the {\em
nonlinear quantum scissors} \cite{Leo97} and have found some
analytical formulas for the probability amplitudes corresponding
to these states. We have shown that starting from the vacuum state
$|0\rangle _a|0\rangle _b$ of our system, its evolution leads to
Bell-like states generation.  Moreover, we have calculated the
concurrence and its behaviour indicates that the ME states are
produced for the system if the photon leakage rates out of the
cavities are less than $10^{-5}$ for the chosen coupling and
nonlinearity parameters. Moreover, we have shown that the
concurrence exhibits some modulation effect as a result of the
existence of various couplings in our system. For each of the
couplings we have some frequency and their interference leads to
some long-frequency oscillations  in the system.

We see, that our model, despite its simplicity, exhibits
intriguing features. We can say that the properties of the system
discussed here are much desired from the point of view of the
physical properties of the nonlinear couplers. Our scheme can be
used for generation of the entangled optical qubits from classical
light, which is a basic but rather simple quantum information
problem. Introduction of conditional measurement in this scheme,
along the lines of Duan et al. proposal \cite{Dua00}, is probably
worth of further study from the point of view of more
sophisticated quantum information applications
\cite{Dua00,Vit00,Ott03}.

Finally, we mention about the experimental feasibility of the
presented scheme. Since our solution is applicable only when the
cavity-field intensities are very small, so an objection arises
that the Kerr nonlinearities are usually negligible in this case.
However, the recent breakthrough advances on nonlinear optics
involving very weak light fields show that the nonlinearities can
be enhanced by several orders of magnitude in ultracold atomic
systems using electromagnetically induced transparency when
resonant optical absorption is eliminated (\cite{Luk01} and
references therein). In particular, giant Kerr nonlinearities have
been theoretically predicted \cite{Sch96} and first experimentally
measured to be $\sim 10^6$ greater than those in the conventional
optical materials \cite{Hau99}. Thus, we believe that the scheme
discussed here can be feasible experimentally.

\section*{Acknowledgments}

Authors wish to thank Profs. Ryszard Tana\'s, Jaros\l{}aw  Zaremba
and Ji\v{r}\'\i{} Bajer for their valuable discussions and
suggestions.


\end{document}